\documentclass[a4paper, pra, twocolumn,superscriptaddress,showpacs,english]{revtex4-1}
\usepackage[T1]{fontenc}
\usepackage[latin9]{inputenc}
\usepackage{float}
\usepackage{amssymb}
\usepackage{amsmath}
\usepackage{graphicx}
\usepackage{babel}
\usepackage{color}

\begin{document}

\title{Fulde-Ferrell-Larkin-Ovchinnikov pairing states between $s$- and $p$-orbital fermions}

\author{Shu-Yang Wang}
\affiliation{State Key Laboratory of Mesoscopic Physics, School of Physics, Peking
University, Beijing 100871, China}
\author{Jing-Wei Jiang}
\affiliation{State Key Laboratory of Mesoscopic Physics, School of Physics, Peking
University, Beijing 100871, China}
\author{Yue-Ran Shi}
\affiliation{Department of Physics, Renmin University of China, Beijing 100872, China}
\author{Qiongyi He}
%\email{qiongyihe@pku.edu.cn}
\affiliation{State Key Laboratory of Mesoscopic Physics, School of Physics, Peking
University, Beijing 100871, China}
\affiliation{Collaborative Innovation Center of Quantum Matter, Beijing 100871, China}
\affiliation{Collaborative Innovation Center of Extreme Optics, Shanxi University, 
Shanxi 030006, China}
\author{Wei Zhang}
\email{wzhangl@ruc.edu.cn}
\affiliation{Department of Physics, Renmin University of China, Beijing 100872, China}
\affiliation{Beijing Key Laboratory of Opto-electronic Functional Materials and Micro-nano Devices,
Renmin University of China, Beijing 100872, China}

\begin{abstract}
We study pairing states in an largely imbalanced two-component Fermi gas loaded in an anisotropic two-dimensional optical lattice, where the spin up and spin down fermions filled to the $s$- and $p_x$-orbital bands, respectively. We show that due to the relative inversion of band structures of the $s$ and $p_x$ orbitals, the system favors pairing between two fermions on the same side of the Brillouin zone, leading to a large stable regime for states with finite center-of-mass momentum, i.e., the  Fulde-Ferrell-Larkin-Ovchinnikov (FFLO) state. In particular, when the two Fermi surfaces are close in momentum space, a nesting effect stabilizes a special kind of $\pi$-FFLO phase with spatial modulation of $\pi$ along the easily tunneled $x$-direction. We map out the zero temperature phase diagrams within mean-field approach for various aspect ratio within the two-dimensional plane, and calculate the Berezinskii-Kosterlitz-Thouless (BKT) transition temperatures $T_{\rm BKT}$ for different phases. 
\end{abstract}
\pacs{67.85.Lm, 03.75.Ss, 05.30.Fk}
\maketitle

%%%%%%%%%%%%%%
%%%%%%%%%%%%%%
\section{Introduction}
\label{sec:intro}

Pairing between fermions residing on separate Fermi surfaces is one of the central questions in the fields of superconductors in a variety of solid state systems~\cite{casalbuoni-04}, color superconductivity in quark matter~\cite{alford-01}, and superfluidity in ultracold atomic gases~\cite{liao-10}. As was first intrigued by the study of magnetic field effect on superconductivity, the discussion on this interesting topic leads to proposals of various exotic pairing states, including the Fulde-Ferrell-Larkin-Ovchinnikov (FFLO)~\cite{flude-64, larkin-65}, breached pair~\cite{liu-03}, and deformed Fermi surface phases~\cite{sarma-63, muther-02}. Among these candidates, the FFLO state consists of pairs formed by two fermions on top of each individual Fermi surface, and is characterized by a finite center-of-mass momentum. Thanks for the high controllability and the separation of charge and spin degrees of freedom, ultracold atomic gases provide a versatile platform to study pairing physics with mismatched Fermi surfaces. A large volume of investigations, both experimental and theoretical, suggest that although the FFLO state may be restricted in a narrow parameter regime for a three-dimensional two-component Fermi gas with mismatched Fermi surfaces~\cite{sheehy-07}, its existence is favored in low dimensions~\cite{orso-07, hu-07} and in systems with synthetic spin-orbit coupling~\cite{zhang-13, yi-review}, where the fluctuations of the order parameter wave vector are restricted by the reduction of symmetry.

In addition to Fermi gases confined in harmonic traps, pairing with mismatched Fermi surfaces are also analyzed theoretically for fermions loaded in optical lattices~\cite{torma-08}. One important finding is that due to the nesting effect between the two Fermi surfaces, the FFLO state can be remarkably favored as the nesting condition is satisfied. This effect is eminent in two-dimensions where a van Hove singularity emerges when the nesting is perfect. Besides, recent studies suggest that when the pairing takes place between fermions residing on different orbital bands, a special $\pi$-FFLO phase with center-of-mass momentum $q = \pi / d$ can be stabilized in a large parameter regime for a quasi-one-dimensional (quasi-1D) lattice potential with lattice spacing $d$~\cite{liu-10, liu-16}. The emergence of such an exotic state is also a result of nesting effect induced by the relative inversion of the single-particle band structures of the two spin components. This $\pi$-FFLO phase, or equivalently referred as $\pi$-phase, has been studied in heterostructures of ferromagnetic and superconducting layers~\cite{buzdin-review}, high $T_c$ supercondcutors~\cite{bernhard-99, maclaughlin-99, chmaisssem-00}, and spin-dependent quasi-1D optical lattices~\cite{demler-10}, and has potential application for quantum computing in building up superconducting qubits via the $\pi$-junctions~\cite{mooij-99, ioffe-99}.
 
In this work, we study inter-band pairing within two-component fermions loaded in a two-dimensional (2D) optical lattice. We show that a $\pi$-FFLO phase can be stabilized due to the nesting effect between the $s$- and $p$-orbital bands. By employing a mean-field approach, we map out the zero temperature phase diagram by varying the chemical potential of each spin component, and find that the FFLO states, either the $\pi$-phase or the conventional FFLO state, is favored within a large parameter window. As a comparison, the Bardeen-Cooper-Schrieffer (BCS) state with zero center-of-mass pairing momentum is not stable under the mean-field level calculation. We also study the evolution of phase diagrams by reducing the hopping integral along one direction, and recover the result for 1D configurations. We then take into account the phase fluctuation on top of the mean-field order parameter, and obtain the Berezinskii-Kosterlitz-Thouless (BKT) transition temperature $T_{\rm BKT}$ for the $\pi$-FFLO, conventional FFLO, and the BCS phases. 

The remainder of this paper is organized as follows. In Sec.~\ref{sec:model}, we present the model under consideration and the mean-field formalism. By minimizing the mean-field thermodynamic potential, we discuss the zero-temperature phase diagrams for various lattice configurations in Sec.~\ref{sec:T0}. We then include the phase fluctuations and obtain the BKT transition temperature in Sec.~\ref{sec:TBKT}. Finally, we summarize the main findings in Sec.~\ref{sec:con}.

%%%%%%%%%%
\section{Model}
\label{sec:model}

We consider a two-component Fermi gas loaded in a quasi-two-dimensional (quasi-2D) cubic optical lattice. The Hamiltonian reads
\begin{eqnarray}
\label{eqn:H1}
H &=& \int d^3 \boldsymbol{r} \sum_{\sigma = \uparrow, \downarrow} \psi_{\sigma}^\dagger (\boldsymbol{r}) 
\left[ - \frac{\hbar^2 }{2m} \nabla^2 + V_{\rm ol}(\boldsymbol{r}) - \mu_\sigma \right] \psi_\sigma(\boldsymbol{r})
\nonumber \\
&& +
g \int d^3 \boldsymbol{r} \psi_{\uparrow}^\dagger (\boldsymbol{r}) \psi_{\downarrow}^\dagger (\boldsymbol{r})
\psi_{\downarrow} (\boldsymbol{r}) \psi_{\uparrow} (\boldsymbol{r}),
\end{eqnarray}
where $\psi_\sigma^\dagger(\boldsymbol{r})$ and $\psi_\sigma(\boldsymbol{r})$ are creation and annihilation operators for fermions at position $\boldsymbol{r}$ and spin $\sigma$, $V_{\rm ol} = \sum_{i = x,y,z} V_{0i} \sin^2(\pi r_i/d)$ is the lattice potential, $\mu_\sigma$ is the chemical potential for spin $\sigma$, and $g$ is the strength for a contact interaction. In the following discussion, we focus on an anisotropic configuration with $V_{0z}$ much greater than $V_{0y}$ and $V_{0x}$, such that the hopping integral along the $z$ direction is negligible to ensure quasi-two-dimensionality. Besides, we also focus on the case with $V_{0y} > V_{0x}$. This in-plane anisotropy breaks the $C_4$ rotational symmetry and lift the degeneracy between the $p_x$ and $p_y$ orbitals, so that we can concentrate on the energetically favorable $p_x$ orbital only. We may then refer the $p_x$ orbital simply as the $p$ orbital to simplify notation unless specified.
\begin{figure}[t]
\centering{}
\includegraphics[width=1.0\columnwidth]{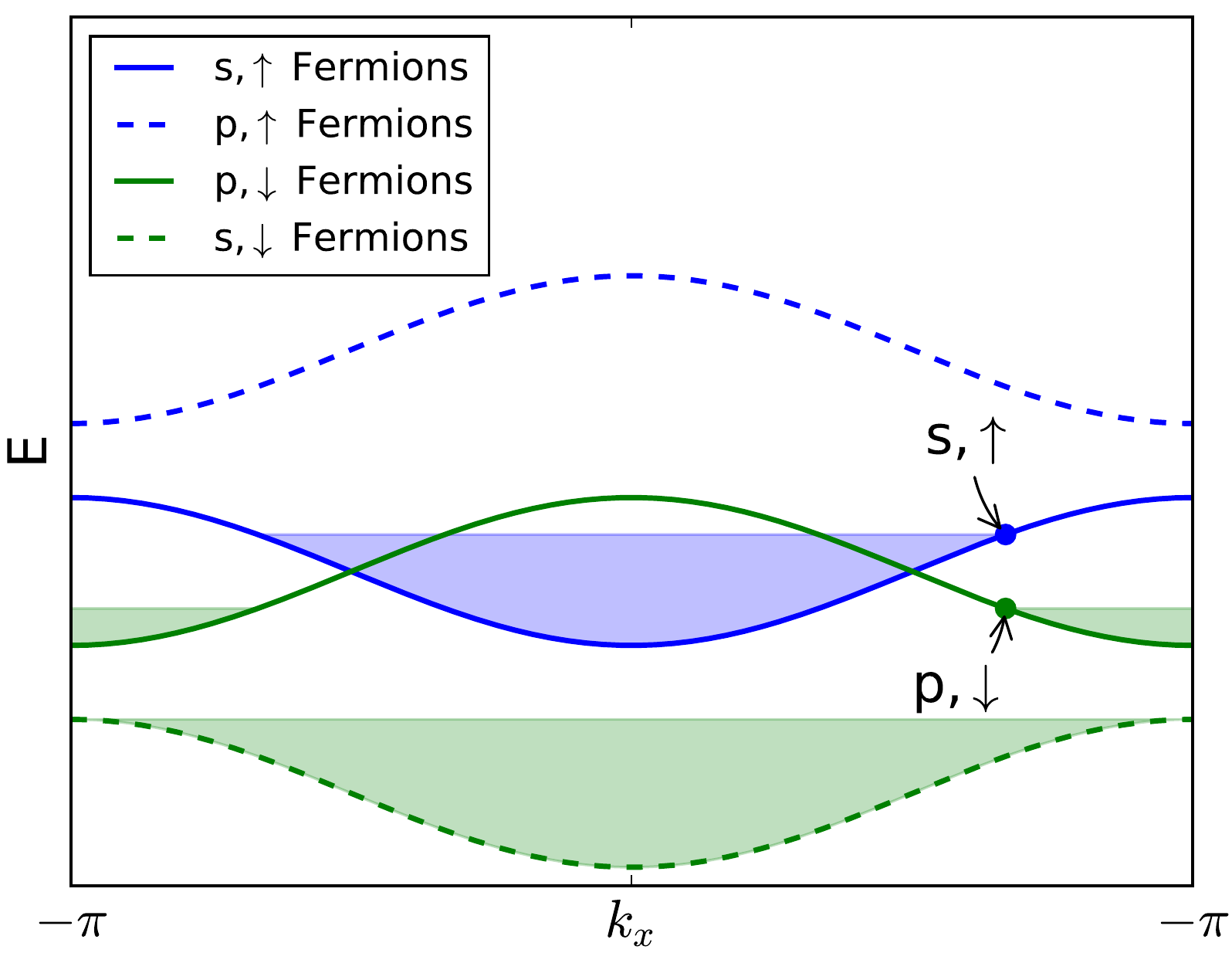}
\caption{(Color online) Schematic illustration of band occupation for spin-up (blue) and spin-down (green) particles. The spin-down fermions in the $s$ band (dashed green) are inert due to a large band gap. Pairing takes place between the spin-up fermions in the $s$ band (solid blue) and the spin-down fermions in the $p$ band (solid green). As the dispersions of the $s$ and $p$ bands have opposite curvatures, the system favors pairing between two particles on the same side of the Brillouin zone, leading to a pairing state with finite center-of-mass momentum.}
\label{fig:scheme}
\end{figure}

Under the condition of a large population imbalance such that the spin-up fermions are filled up to the $s$ orbital, while the spin-down particles are filled to the $p$ band, the pairing will take place between particles residing on the Fermi surfaces in the $s$ and $p$ bands, as schematically illustrated in Fig.~\ref{fig:scheme}. A minimal 2D Hamiltonian involves only these two relevant bands, and takes the following form under the tight-binding approximation
\begin{eqnarray}
\label{eqn:Htb}
H_{\rm 2D} &=& \sum_{{\bf k},\sigma} ( \epsilon_{{\bf k},\sigma} - \mu_\sigma) c_{{\bf k},\sigma}^\dagger c_{{\bf k},\sigma}
\nonumber \\
&& +
U \sum_{{\bf k},{\bf k}^\prime,{\bf q}} 
c_{{\bf k}+{\bf q},\uparrow}^\dagger c_{-{\bf k}+{\bf q},\downarrow}^\dagger 
c_{-{\bf k}^\prime+{\bf q},\downarrow} c_{{\bf k}^\prime+{\bf q},\uparrow}.
\end{eqnarray}
Here, we have integrated out the degrees of freedom along the strongly confined $z$-direction. The single-particle dispersion reads 
\begin{eqnarray}
\label{eqn:dispersion}
\epsilon_{{\bf k},\sigma} = 2 J_{x \sigma} (1-\cos k_x) + 2 J_{y \sigma} (1-\cos k_y)
\end{eqnarray}
with hopping integrals $J_{x\uparrow}, J_{y\uparrow}, J_{y\downarrow} >0$, and $J_{x\downarrow} <0$, reflecting the symmetries of $s$- and $p_x$-orbitals for spin-up and spin-down particles, respectively. The specific values of hopping coefficients are determined by the overlap of Wannier functions of corresponding bands at adjacent sites. The on-site interaction $U$ is also obtained by the density-density overlap of on-site Wannier functions, and can be tuned by either changing the contact interaction strength $g$ via a Feshbach resonance, or by varying the $z$-direction lattice depth through a confinement-induced resonance~\cite{petrov-01, kestner-07, zhang-08}.

By defining a pairing order parameter $\Delta_{2 \bf q} \equiv U \sum_{\bf k} c_{-{\bf k}+{\bf q},\downarrow} c_{{\bf k}+{\bf q},\uparrow}$, we obtain the mean-field Hamiltonian
\begin{eqnarray}
\label{eqn:Hmf}
H_{\rm MF} &=& - \frac{|\Delta_{2{\bf q}}|^2}{U} + \sum_{\bf k} \Big( \xi_{{\bf k},\uparrow} c_{{\bf k},\uparrow}^\dagger c_{{\bf k},\uparrow}
+  \xi_{{\bf k},\uparrow} c_{{\bf k},\uparrow}^\dagger c_{{\bf k},\uparrow}
\nonumber \\
&& + \Delta_{2{\bf q}} c_{{\bf k}+{\bf q},\uparrow}^\dagger c_{-{\bf k}+{\bf q},\downarrow}^\dagger 
+  \Delta_{2{\bf q}}^\dagger c_{-{\bf k}+{\bf q},\downarrow} c_{{\bf k}+{\bf q},\uparrow}
\Big),
\end{eqnarray}
where $\xi_{{\bf k},\sigma} = \epsilon_{{\bf k},\sigma} - \mu_\sigma$. Integrating out the fermionic degrees of freedom, we obtain the thermodynamic potential,
\begin{eqnarray}
\label{eqn:omega}
\Omega &=& - \frac{|\Delta_{2\bf q}|^2}{U} + \sum_{\bf k} \Big[
\xi_{-{\bf k}+{\bf q}, \downarrow} 
\nonumber \\
&& 
+ \frac{1}{\beta} \ln \left( (1+ e^{-\beta E_{{\bf k},{\bf q},+}}) (1+e^{-\beta E_{{\bf k},{\bf q},-}})\right)
\Big],
\end{eqnarray}
where $\beta = 1/k_B T$ is the inverse temperature, and the two branches of quasi-particle dispersions are given by
\begin{eqnarray}
\label{eqn:qparticle}
E_{{\bf k},{\bf q},\pm} &=& \frac{\xi_{{\bf k}+{\bf q},\uparrow} - \xi_{-{\bf k}+{\bf q},\downarrow}}{2} 
\nonumber \\
&& \pm
\sqrt{\left(\frac{\xi_{{\bf k}+{\bf q},\uparrow} +  \xi_{-{\bf k}+{\bf q},\downarrow}}{2}\right) 
+ |\Delta_{2\bf q}|^2}.
\end{eqnarray}
The ground state of system can then be determined by minimizing the thermodynamic potential by varying the amplitude $\Delta \equiv |\Delta_{2\bf q}|$ and the wave vector ${\bf q}$ of the order parameter.

%%%%%%%%%%%%
\section{Zero temperature phase diagrams}
\label{sec:T0}
\begin{figure}[t]
\centering{}
\includegraphics[width=0.98\columnwidth]{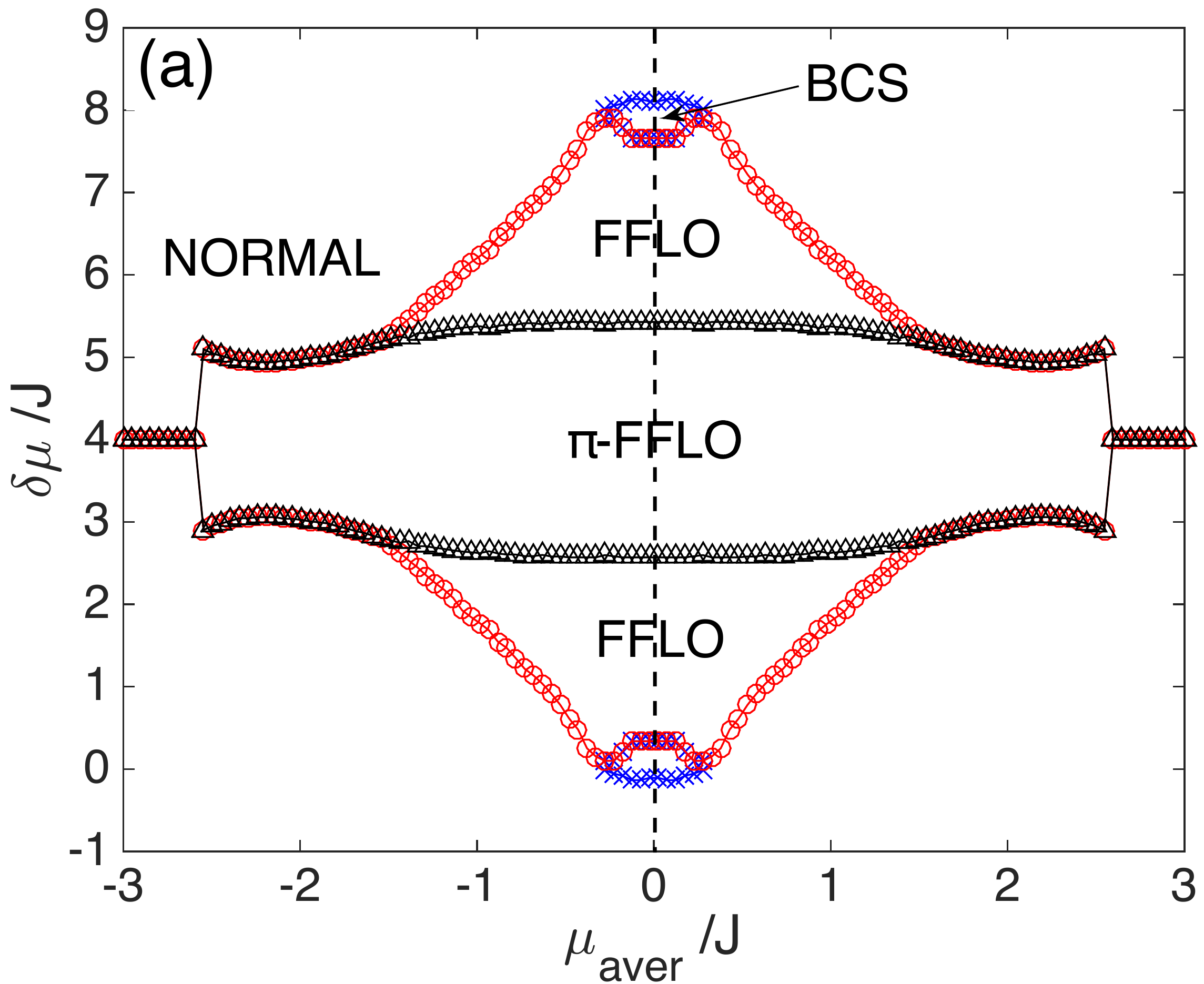}
\includegraphics[width=0.99\columnwidth]{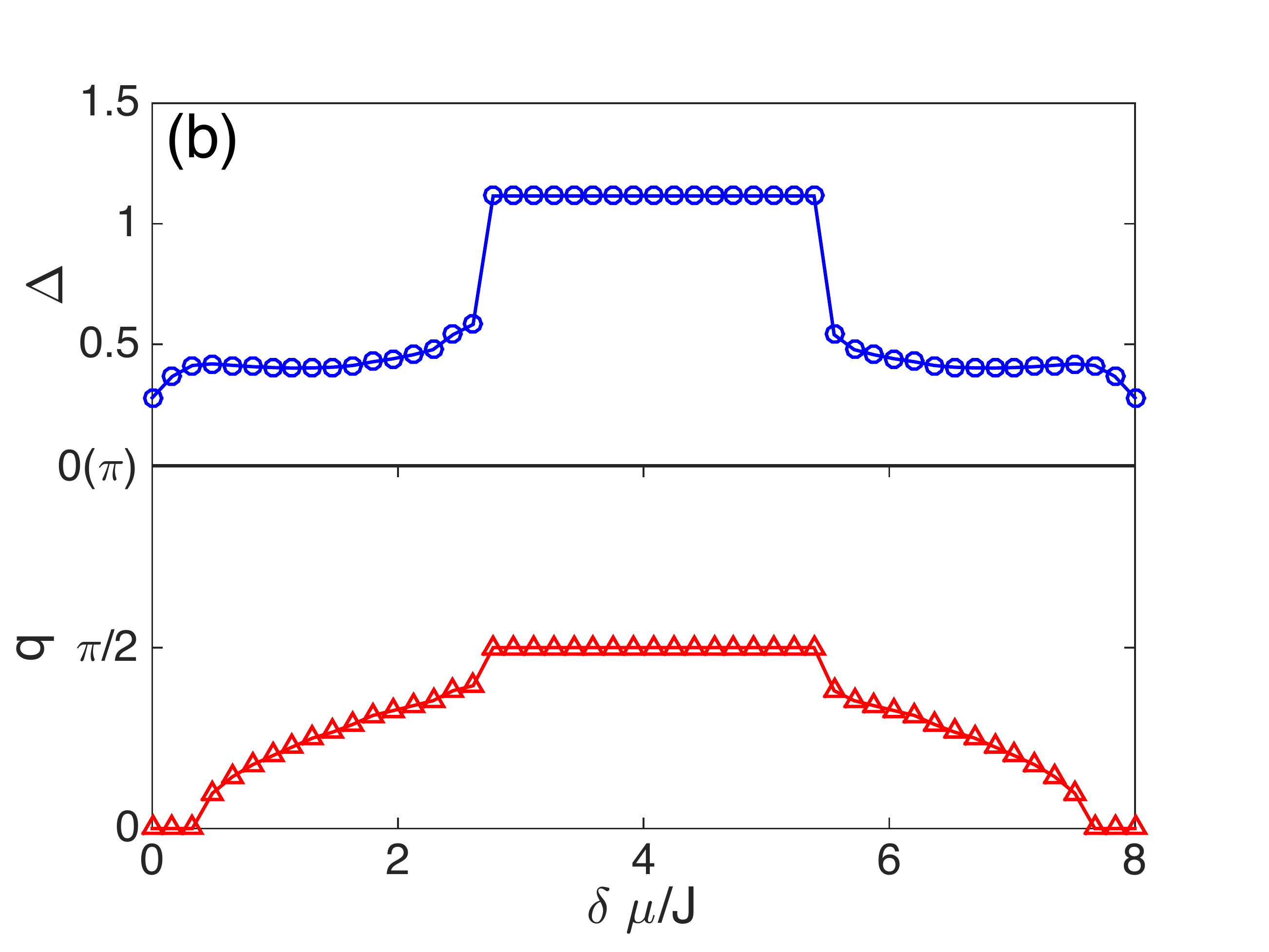}
\caption{(Color online) (a) The phase diagram of one-dimensional configuration with $J_{y\sigma}/J = 0$ by varying the average chemical potential $\mu_{\rm aver}$ and chemical potential difference $\delta \mu$. The $\pi$-FFLO phase is characterized by a center-of-mass momentum of $q d = \pi/2$. (b) The variations of $\Delta$ and $q$ by changing $\delta \mu$ with $\mu_{\rm aver}/J = 0$, as moving along the dashed line on the phase diagram (a). The $\pi$-FFLO--FFLO and the BCS--Normal phase transitions are of the first order, while the FFLO--BCS transition is of the second order. Other parameters used here are $J_{x \uparrow}/J = 1$, $J_{x \downarrow}/J = -1$, and $U/J = -3.3$.}
%{\color{red} CHANGE (A): $\delta \mu \in [-4,4]$, $\mu_{\rm aver} \in [0, 4]$. CHANGE (B): $\delta \mu \in [-4,4]$. CHANGE (B): $\Delta /J$, $2q d$, ONLY ONE BRANCH OF $q$ FOR FFLO PHASE.}}
\label{fig:1D}
\end{figure}

We first consider the 1D case with $J_{y\sigma} = 0$. This configuration has been analyzed in previous works using mean-field approach and density matrix renormalization group method~\cite{liu-10, liu-16}, which both suggest large parameter windows for the $\pi$-FFLO and the conventional FFLO states. A brief discussion of this limiting case and a comparison with existing results can be useful to explain our methods and findings, and to describe the physics behind.

In Fig.~\ref{fig:1D}(a), we show the mean-field phase diagram by varying the chemical potentials of the two spin species. We choose the hopping integral for spin-up along the $x$-direction to be the energy unit $J_{x\uparrow} = J = 1$, and set $J_{x\downarrow} = -1$ to accommodate the symmetry of the $p$-orbital band. In Fig.~\ref{fig:1D}, we define the average chemical potential $\mu_{\rm aver}$ and the chemical potential difference $\delta \mu$ as
\begin{eqnarray}
\label{eqn:mu}
\mu_{\rm aver} &=& \frac{\mu_{\uparrow} + \mu_{\downarrow}}{2},
\nonumber \\
\delta \mu &=& \mu_\uparrow - \mu_\downarrow.
\end{eqnarray}
Notice that the single-particle dispersions for spin-up and spin-down fermions range within $\epsilon_{{\bf k},\uparrow}/J \in [0,4]$ and $\epsilon_{{\bf k},\downarrow}/J \in [-4,0]$, respectively. As the $s$- and $p$-orbital bands are symmetric under reflection for this special choice of parameters, the phase diagram is also symmetric along $\mu_{\rm aver}/J = 0$ and $\delta \mu/J = 4$.

One striking feature of this phase diagram is a large stable region for the $\pi$-FFLO state, which is characterized by a finite order parameter amplitude $\Delta$ and a center-of-mass momentum $2 q d/\pi$, as illustrated in Fig.~\ref{fig:1D}(b). This $\pi$-FFLO phase features a center-of-mass $p$-wave symmetry, hence is another example of $p$-wave superfluids induced by the orbital degrees of freedom~\cite{wu-11, liu-14, liu-16}. Notice that the $\pi$-FFLO phase presents near the regime $\delta \mu/J \sim 4$, where the Fermi surfaces for the two spin species are close in the first Brillouin zone. This observation is consistent with the understanding of the pairing mechanism for the $\pi$-FFLO state, i.e., the nesting effect between the two Fermi surfaces connected by $2 \pi / d$. When the chemical potential difference $\delta \mu/J$ is moved further away from 4, the nesting condition for the two Fermi surfaces is no longer a momentum shift of $2 \pi /d$. In this case, the $\pi$-FFLO becomes energetically unfavorable than a conventional FFLO state which processes the correct center-of-momentum $2q$ for nesting. The transition between $\pi$-FFLO and conventional FFLO is of the first order, as shown in Fig.~\ref{fig:1D}(b). Finally, We also identify a small BCS regime around $\mu_{\rm aver}/J \approx 0$ and $\delta \mu/J \approx 0$ or $8$. These regimes correspond to the condition that the band of one spin species is almost completely filled while the other is nearly empty, so that the pairing takes place between two fermions either residing near the center or the opposite boundaries of the first Brillouin zone, leading to a pairing state with zero center-of-mass momentum. The BCS-FFLO phase boundary is of the second order, as suggested by the smooth variations of $\Delta$ and $q$ shown in Fig.~\ref{fig:1D}(b).
\begin{figure}[t]
\centering{}
%\includegraphics[width=0.48\columnwidth]{fig3a-2ddiagram.pdf}
%\includegraphics[width=0.48\columnwidth]{fig3b-2ddeltaq.pdf}
%\\
\includegraphics[width=0.98\columnwidth]{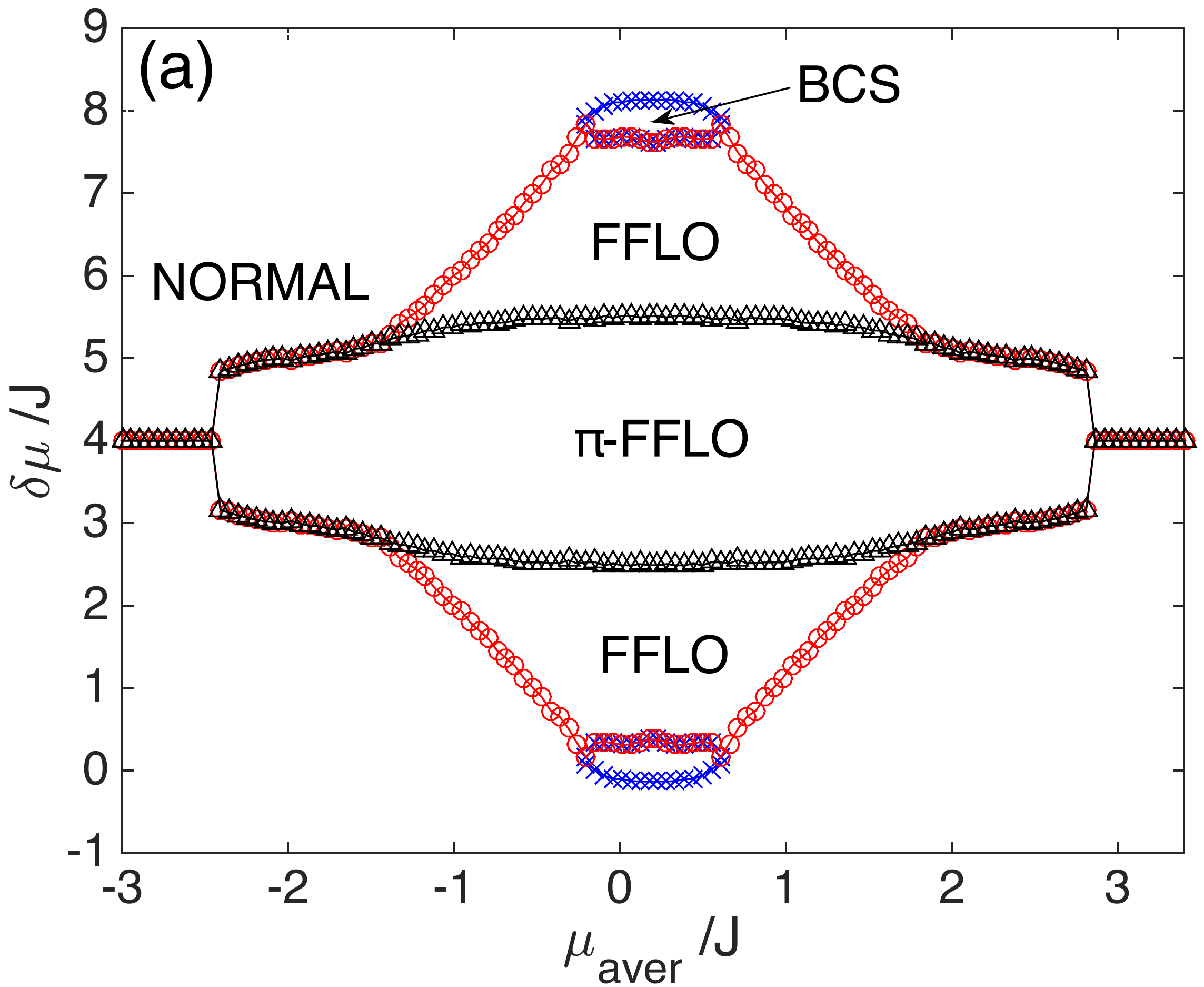}
\includegraphics[width=0.98\columnwidth]{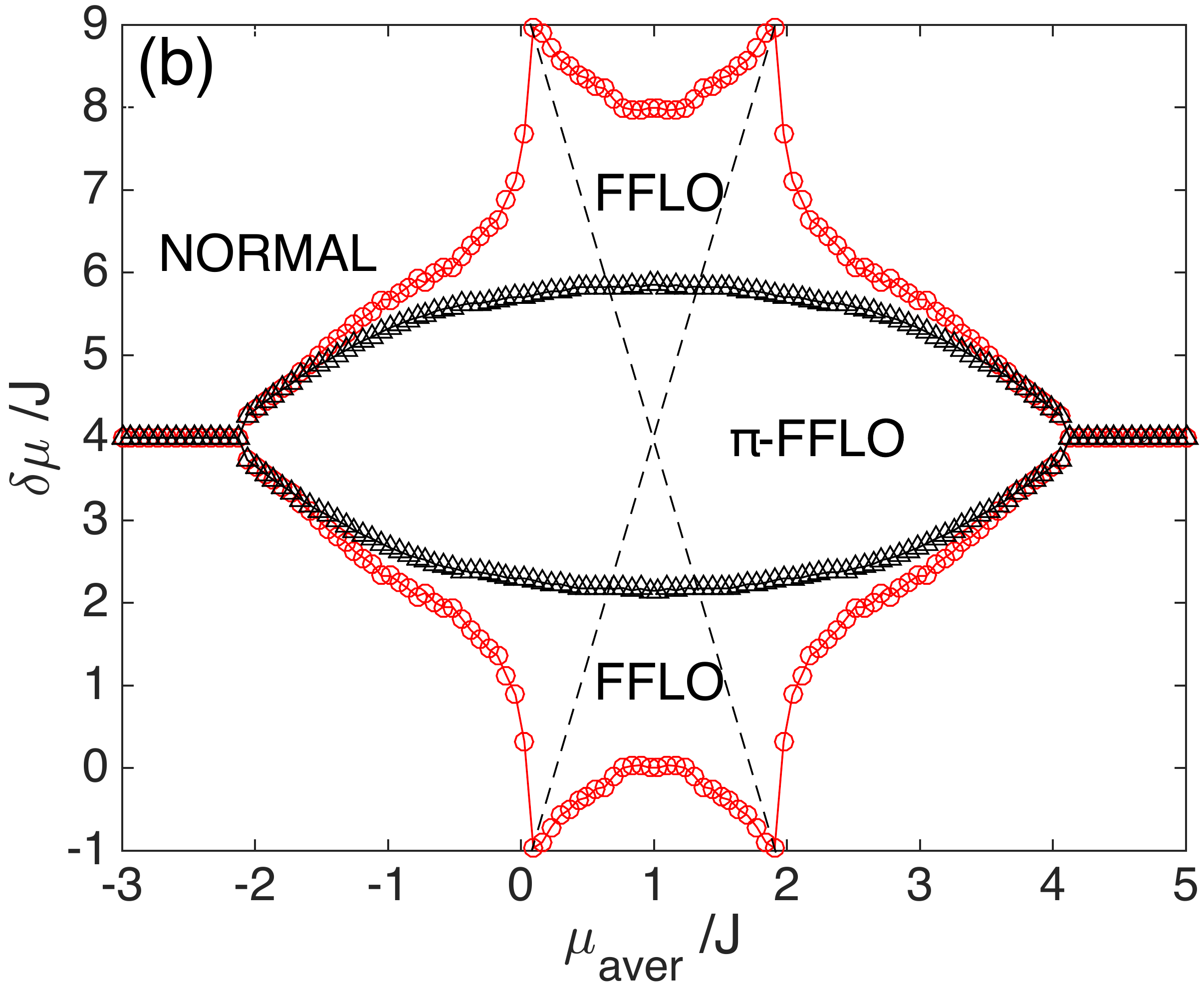}
\caption{(Color online) The phase diagram for a two-dimensional system with (a) $J_{y \sigma}/J = 0.1$ and (b) $J_{y \sigma}/J = 0.5$. Other parameters are the same as in Fig.~\ref{fig:1D}.}
% {\color{red} CHANGE (A) $\mu_{\rm aver} \in [-2, 2.2]$. CHANGE (B) $\mu_{\rm aver} \in [-2, 3]$.}
\label{fig:2D}
\end{figure}

With the understanding of the 1D phase diagram and the properties of various phases therein, next we discuss the 2D configuration with $J_{y\sigma}/J \neq 0$. Remind that we consider only the anisotropic 2D lattice with the lattice potential along the $y$-direction is higher than that along the $x$-direction, such that we only need to take the lowest lying $p_x$ orbital into consideration. In Fig.~\ref{fig:2D}(a) and \ref{fig:2D}(b), we show the zero-temperature phase diagrams for $J_{y \sigma}/J = 0.1$ and $0.5$, respectively. Due to the reflection symmetry of the dispersions for spin-up and spin-down particles, the phase diagrams are symmetric along $\delta \mu/J = 4$ and $\mu_{\rm aver} /J = 0.2$ [for Fig.~\ref{fig:2D}(a)] or $\mu_{\rm aver}/J = 1$ [for Fig.~\ref{fig:2D}(b)].

From Fig.~\ref{fig:2D}, we find that the general structure of the phase diagrams is similar to that in 1D. Specifically, there is a large region for stable $\pi$-FFLO around $\delta \mu /J \sim 4$. The center-of-mass wave vector for this state is $2 {\bf q} = (\pi / d, 0)$, which indicates a $p_x$-wave symmetry. The $\pi$-FFLO phase is surrounded by a conventional FFLO phase characterized by a wave vector $2 {\bf q} = (2 q_x, 0)$ via a first-order phase transition.

There are, however, some distinct features present in the 2D case, in particular with large $J_{y \sigma}/J$ as shown in Fig.~\ref{fig:2D}(b). First, the stable region for the FFLO state is significantly extended when the chemical potentials satisfy a certain condition, as depicted by dotted lines in Fig.~\ref{fig:2D}(b). Along those lines, the two Fermi surfaces of different spin species can have a perfect nesting, leading to a van Hove singularity with logarithmic diverging density of states and hence a pairing instability. Second, the BCS state disappears with increasing $J_{y \sigma}/J$. This is also a consequence of the enhanced FFLO instability induced by the 2D nesting condition.

%%%%%%%%%
\section{Superfluid transition temperature}
\label{sec:TBKT}

In this section, we go beyond the mean-field approximation by considering $\Delta_{2 \bf q} = |\Delta| + |\delta \Delta| e^{i \theta}$, where $|\delta \Delta|$ and $\theta$ are the amplitude and phase of the fluctuations atop the mean-field saddle point solution~\cite{carlos-06}. By substituting the expression  above to the thermodynamic potential Eq.~\ref{eqn:omega} and integrating out the amplitudes, we obtain the superfluid density
\begin{eqnarray}
\label{eqn:SFden}
\rho_{ij} = \frac{m}{\hbar^2} \left( \frac{\partial^2 \Omega}{\partial q_{si} \partial q_{sj}}\right) \bigg \vert_{{\bf q}_s=0},
\end{eqnarray} 
where ${\bf q}_s = (q_{sx}, q_{sy}) \equiv (\partial_x \theta, \partial_y \theta)$ is the wave vector associated with the superfluid velocity. Notice that the superfluid density is a tensor for the most general case with anisotropy. The diagonal elements read
\begin{eqnarray}
\label{eqn:J}
\rho_{xx} &=& \frac{m}{\hbar^2} \sum_{{\bf k}, \eta=\pm} 
\Bigg[
\left(c_- + \eta \frac{\xi c_+ + \frac{\Delta^2}{\xi^2 + \Delta^2} s_+^2}{\sqrt{\xi^2 + \Delta^2}} \right) f(E_\eta)
\nonumber \\
&& - \frac{\beta}{4}
\left( s_- + \eta \frac{\xi}{\sqrt{\xi^2 + \Delta^2}} s_+ \right)^2 {\rm sech}^2\frac{\beta E_\eta}{2}
\Bigg],
\nonumber \\
\rho_{yy} &=& \frac{m}{\hbar^2} \sum_{{\bf k}, \eta=\pm} 
\Bigg[
\eta \frac{2 J_{y} \xi \cos k_y}{\sqrt{\xi^2 + \Delta^2}} f(E_\eta) 
\nonumber \\
&& - \beta J_y^2 \sin^2 k_y {\rm sech}^2 
\frac{\beta E_\eta}{2}
\Bigg].
\end{eqnarray}
In the expressions above, $f(x) = 1/(1+e^\beta x)$ is the Fermi distribution function, $\xi = (\xi_{{\bf k}+{\bf q}, \uparrow} + \xi_{-{\bf k}+{\bf q}, \downarrow})/2$, and
\begin{eqnarray}
\label{eqn:parameters}
s_\pm &=& J_{x\uparrow} \sin(k_x + q_x) \pm J_{x\downarrow} \sin(-k_x + q_x),
\nonumber \\
c_\pm &=& J_{x\uparrow} \cos(k_x + q_x) \pm J_{x\downarrow} \sin(-k_x + q_x).
\end{eqnarray}
In two dimensions, the superfluid transition temperature is of the Berezinskii-Kosterlitz-Thouless type~\cite{berezinskii-71, kt-72}, which is associated with the association and dissociation of vortex and anti-vortex pairs. The equation for the BKT temperature $T_c = T_{\rm BKT}$ is determined by the Kosterlitz-Thouless condition~\cite{kt-72}
\begin{eqnarray}
\label{eqn:TBKT}
T_{\rm BKT} = \frac{\pi}{2} \sqrt{\rho_{xx} \rho_{yy}}.
\end{eqnarray}
This equation must be solved self-consistently with the minimization condition for the thermodynamic potential Eq. (\ref{eqn:omega}) to determine $\Delta$, ${\bf q}$, and $T_{\rm BKT}$.

\begin{figure}[t]
\centering{}
\includegraphics[width=0.98\columnwidth]{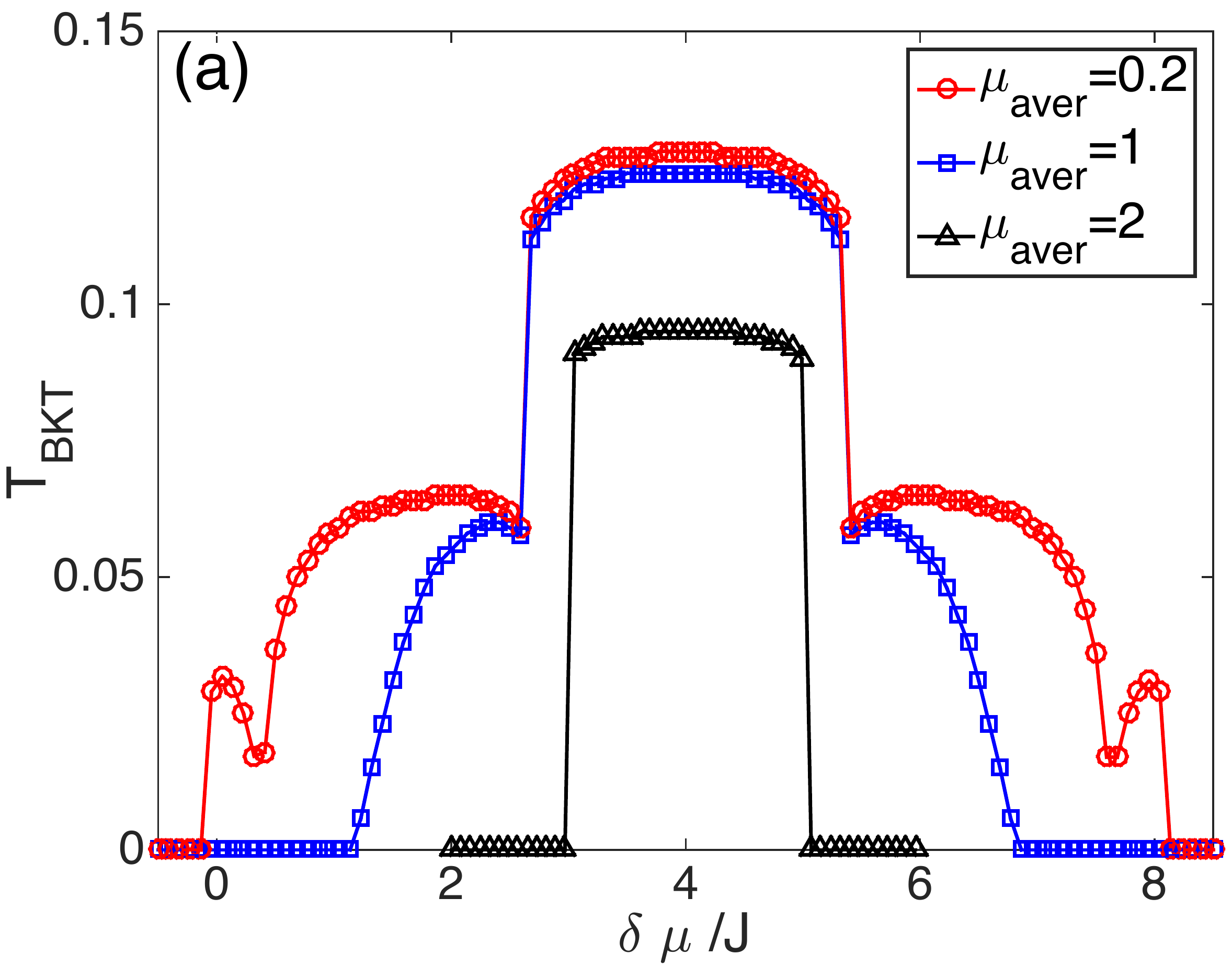}
\includegraphics[width=0.98\columnwidth]{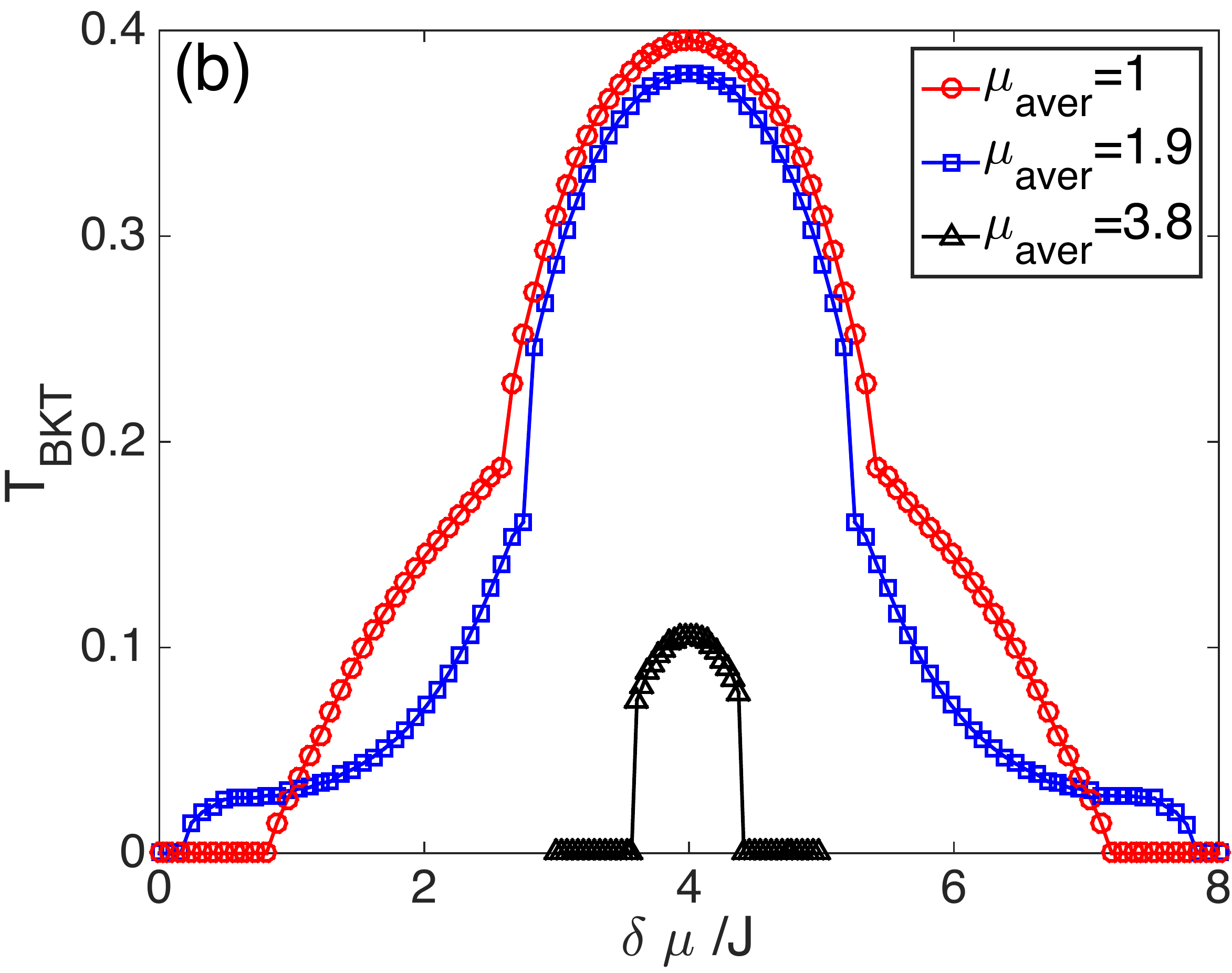}
\caption{(Color online) The superfluid transition temperature $T_{\rm BKT}$ for (a) $J_{y\sigma}/J = 0.5$ and (b) $J_{y\sigma}/J = 0.1$.  Notice that the BCS state acquires an elevated transition temperature due to the increase of superfluid density. Other parameters are the same as in Fig.~\ref{fig:1D}. }
\label{fig:TBKT}
\end{figure}

The solutions for transition temperature $T_{\rm BKT}$ are shown in Fig.~\ref{fig:TBKT}. For the strongly anisotropic case with $J_{y\sigma}/J = 0.1$ as shown in Fig.~\ref{fig:TBKT}(a), the critical temperature reaches its maximum at the center of the $\pi$-FFLO regime, where the two bands are both half filled such that the nesting is perfect. As the two Fermi surfaces deviate from this optimistic condition, $T_{\rm BKT}$ reduces to lower values, and features an abrupt drop when entering the FFLO regime. Interestingly, the BCS state located in the region with most separated Fermi surfaces acquires a slightly elevated critical temperature. As the system crosses from 1D to 2D with $J_{y\sigma}/J = 0.5$ [Fig.~\ref{fig:TBKT}(b)], the highest transition temperature is also achieved when the two bands are both half filled. We also find that $T_{\rm BKT}$ is elevated when the 2D nesting condition is satisfied, as can be seen by comparing results for $\mu_{\rm aver}/J = 1$ and $\mu_{\rm aver}/J = 1.9$ at large chemical potential differences.

%%%%%%%%%%%%%
\section{Conclusion}
\label{sec:con}
We investigate inter-band pairing in an anisotropic two-dimensional optical lattice with the Fermi surface of one spin species locates in the $s$-band and the other in the $p_x$-orbital band. By mapping out the zero-temperature phase diagrams, we conclude that the FFLO pairing states with finite center-of-mass momentum are favored in a large parameter window due to the relative inversion of band structures and the nesting effect. Specifically, a $\pi$-FFLO state with spatial modulation of double lattice spacing along the easily tunneled $x$-direction can be stabilized when the Fermi surfaces for the two spin species are close within the Brillouin zone. We further discuss the fluctuation effect at finite temperatures, and calculate the BKT transition temperature for various phases.

\acknowledgments
This work is supported by the National Natural Science Foundation of China (Grant Nos. 11274009, 11274025, 11434011, 11522436, 11622428, 61475006), National Key R\&D Program (Grant Nos. 2013CB922000, 2016YFA0301201),  and the Research Funds of Renmin University of China (Grant Nos. 10XNL016, 16XNLQ03).

%%%%%%%%%%%

\end{document}